\documentclass[preprint,3p,twocolumn]{elsarticle}

\usepackage{amssymb}
\usepackage{amsmath}

\usepackage[switch, pagewise]{lineno}

\usepackage{graphicx}

\usepackage[]{algorithm2e}

\usepackage{hyperref}

\journal{Applied Mathematics and Computation}

\begin{document}


\newcommand{\myfigure}[3] {
    \begin{figure}
        \centering
        \includegraphics[width=1.0\columnwidth]{#1}
        \caption{#2}
        \label{#3}
    \end{figure}
}

\newcommand{\vect}[1]{\boldsymbol{#1}}

\begin{frontmatter}


\title{A Novel Point Inclusion Test for Convex Polygons Based on Voronoi
Tessellations}

\author[1]{Rahman Salim Zengin
\corref{cor1}
\fnref{fn1}}
\ead{rszengin@itu.edu.tr}

\author[2]{Volkan Sezer
\fnref{fn2}}
\ead{sezerv@itu.edu.tr}

\cortext[cor1]{Corresponding author}

\fntext[fn1]{\href{https://orcid.org/0000-0002-3104-4677}{ORCID: https://orcid.org/0000-0002-3104-4677}}

\fntext[fn2]{\href{https://orcid.org/0000-0001-9658-2153}{ORCID: https://orcid.org/0000-0001-9658-2153}}

\address[1]{Department of Mechatronics Engineering, Istanbul Technical University, Istanbul, Turkey}
\address[2]{Department of Control and Automation Engineering, Istanbul Technical University, Istanbul, Turkey}

\begin{abstract}
    The point inclusion tests for polygons, in other words the point-in-polygon
    (PIP) algorithms, are fundamental tools for many scientific fields related
    to computational geometry, and they have been studied for a long time. The
    PIP algorithms get direct or indirect geometric definition of a polygonal
    entity, and validate its containment of a given point. The PIP algorithms,
    which are working directly on the geometric entities, derive linear boundary
    definitions for the edges of the polygons. Moreover, almost all direct test
    methods rely on the two-point form of the line equation to partition the
    space into half-spaces. Voronoi tessellations use an alternate approach for
    half-space partitioning. Instead of line equation, distance comparison
    between generator points is used to accomplish the same task. Voronoi
    tessellations consist of convex polygons, which are defined between
    generator points. Therefore, Voronoi tessellations have become an
    inspiration for us to develop a new approach of the PIP testing, specialized
    for convex polygons. The equations, essential to the conversion of a convex
    polygon to a Voronoi polygon, are derived. As a reference, a very standard
    convex PIP testing algorithm, \textit{the sign of offset}, is selected for
    comparison. For generalization of the comparisons, \textit{the ray crossing}
    algorithm is used as another reference. All algorithms are implemented as
    vector and matrix operations without any branching. This enabled us to
    benefit from the CPU optimizations of the underlying linear algebra
    libraries. Experimentation showed that, our proposed algorithm can have
    comparable performance characteristics with the reference algorithms.
    Moreover, it has simplicity, both from a geometric representation and the
    mental model.
\end{abstract}


\end{frontmatter}

\section{Introduction}
\label{introduction}

Various point inclusion tests \cite{haines_point_1994} are used in many
applications \cite{alciatore_winding_1995}, including planning for autonomous
driving \cite{ziegler_fast_2010}, geographical information systems
\cite{peucker_cartographic_1975, nordbeck_computer_1967,
longley_geographical_2005}, and computer graphics
\cite{glassner_introduction_1989}. Any improvements on the efficiency of the
point inclusion tests will provide a direct benefit to the mentioned areas.

When autonomous driving related planning applications are considered, planning
is mostly done in a 2D space. Collected real-time sensor data, especially
Lidar-based point cloud data, is mapped to the 2D space. Collision check is one
of the most critical components of the motion planning. Several simplifications
on collected data and vehicle representation is required to make it efficient.  
Modeling the vehicle as a circle or combination of several circles is one of the
widely used techniques for collision check. Although this simplification works
well for most situations, there is always an accuracy problem depending on the
number of circles, that are used to model the vehicle \cite{ziegler_fast_2010}.

In order to make a more accurate collision check, footprint of the car can be
modeled as a convex polygon. In order to make a real-time motion planning,
efficient collision-check algorithms, that are capable of testing big batches of
points against the convex polygon model of the car, are needed. Even though
there are simple and well known algorithms, we propose an alternative algorithm
based on Voronoi approach to accomplish the same task.

Geographical information systems \cite{peucker_cartographic_1975,
nordbeck_computer_1967, longley_geographical_2005} is another field that relies
on point inclusion tests. It is used to process large databases of cartographic
data. Measurements taken in the field must be matched with the prior information
related to the area. Using the measurements, point inclusion tests are run
against big databases to accomplish that task.

Another field, in which the point-in-polygon queries are actively being used, is
computer graphics \cite{glassner_introduction_1989}. A scene contains many
object models, which are composed of polygons. For visualization on the screen,
proper rasterization of the geometric data is needed. To match the pixels on the
screen with the geometric data, the polygons are mapped to the screen plane. Then
membership of every screen pixel is determined via point inclusion testing, so
that the pixels can be painted properly.

Voronoi tessellations consist of convex polygons, and there is a huge literature
related to Voronoi tessellations. Simplest point inclusion tests are based on
line equations and point-to-line distance calculations. Conversion of a convex
polygon to a Voronoi polygon has the advantage of using only point-to-point
distance calculations. Required equations for the conversion of a convex polygon
to a Voronoi polygon are derived step-by-step, throughout this paper.

For completeness, two simple and well known point inclusion
methods are summarized, and then compared with our proposed method. In order
to compare the algorithms in an equal manner, all algorithms are implemented
using vector and matrix operations instead of simple loops, with the help of the
related libraries. In this way, computations are handled more efficiently. As a
result, the proposed algorithm showed comparable performance with the reference
algorithms.

The structure of the paper is as follows: In (Section \ref{background}) two
reference algorithms are mentioned and the notation is given. In (Section
\ref{conversion}), conversion of a convex polygon to a Voronoi polygon is
described, and required equations are derived. In (Section \ref{pip_test}), the
point inclusion testing via the generators is described. In (Section
\ref{analysis}), expected performance of our proposed algorithm is discussed. In
(Section \ref{correctness}), for a certain generated test data, correctness of
our proposed algorithm is proven via comparison against \textit{the sign of
offset} algorithm. In (Section \ref{performance}), experimental setup is
described, experimental results are shared and discussed.

\section{Background}
\label{background}

The reference point inclusion algorithms are explained. Then, notation of the
paper is given.

\subsection{The ray crossing method}
\label{ray_crossing}

\begin{figure}
    \centering
    \includegraphics[width=0.85\columnwidth]{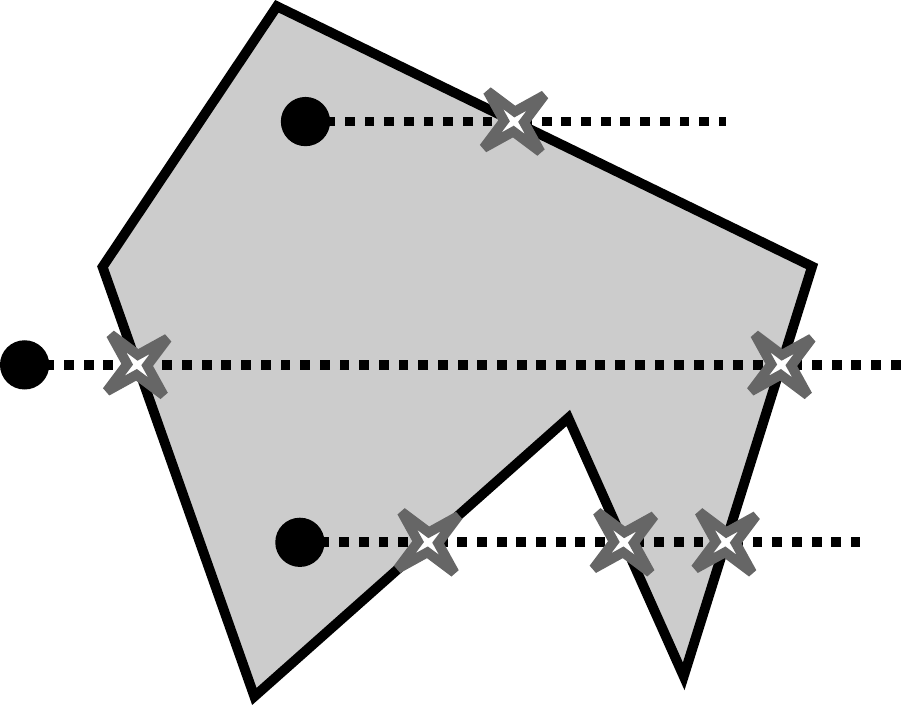}
    \caption{Ray crossing method}
    \label{fig:ray_crossing}
\end{figure}

\textit{The ray crossing} method
\cite{shimrat_algorithm_1962,haines_point_1994,franklin_pnpoly_2020} is the
golden standard of the point inclusion tests. It can be used for simple
polygons. As shown in the (Figure \ref{fig:ray_crossing}) a ray directed to the
$+x$ direction is used to count crossings of the ray and the polygon. If the ray
crosses the polygon edges in odd numbers it is inside, otherwise it is outside.

All edges of the polygon are checked whether they are on the same $y$ level of
the point. If applicable, line equation in the two-point form
\cite{weisstein_two-point_nodate} is used to determine the half-plane of the
point. For a $+x$ going ray it must be on the left half-plane. If so, it is
counted as a crossing.

The pseudocode of the ray crossing implementation, which is used for
experiments, is given in (Algorithm \ref{alg:crossing}).

\begin{algorithm}
    \DontPrintSemicolon
    \SetKwProg{Fn}{Function}{}{}
    \SetKwFunction{Crossing}{CrossingInclusion}
    \Fn(){\Crossing}{}{
        \KwData{\\
            \tcp{$V$: Vertices}
            $V \longleftarrow \begin{bmatrix}
                \vect{v_1} & \cdots & \vect{v_n}
            \end{bmatrix}$\;
            \tcp{$Q$: Query Points}
            $Q \longleftarrow \begin{bmatrix}
                \vect{q_1} & \cdots & \vect{q_m}
            \end{bmatrix}$\;
        }
        \KwResult{$IsIn$: Boolean}
        \Begin{
            \tcp{$V'$: Rolled Vertices}
            $V' \longleftarrow \begin{bmatrix}
                \vect{v_n}, \vect{v_1} & \cdots & \vect{v_{n-1}}
            \end{bmatrix}$\;
            \tcp{$\Delta$ of successive vertices}
            $\Delta V \longleftarrow (V - V')$\;
            \tcp{Edges in y range}
            $InRange \longleftarrow (V_y > Q_y) \oplus (V'_y > Q_y)$\;
            \tcp{Is edge going up?}
            $GoingUp \longleftarrow V_y > V'_y$\;
            \tcp{LHS \& RHS of the line equation}
            $LHS \longleftarrow Q_y \circ \Delta V_x - Q_x \circ \Delta V_y$\;
            $RHS \longleftarrow V'_y \circ \Delta V_x - V'_x \circ \Delta V _y$\;
            \tcp{Is point on the left}
            $OnLeft \longleftarrow GoingUp$ ?
                $(LHS > RHS)$ :
                $(LHS < RHS)$\;
            $Crossing \longleftarrow InRange \wedge OnLeft$
            $IsIn \longleftarrow Mod_2(\sum_i Crossing_{ij}) = 0$
        }
    }
    \caption{The ray crossing point inclusion test}
    \label{alg:crossing}
\end{algorithm}

\subsection{The sign of offset method}
\label{sign_of_offset}

\begin{figure}
    \centering
    \includegraphics[width=0.85\columnwidth]{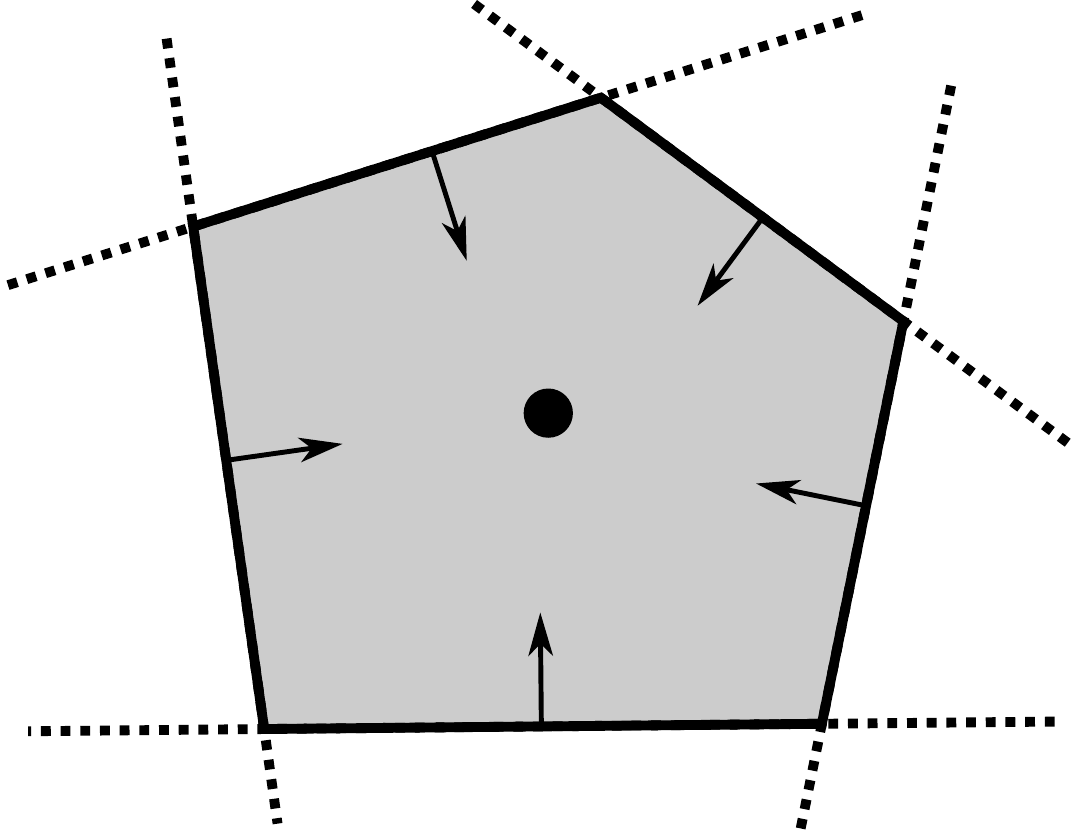}
    \caption{The sign of offset method}
    \label{fig:sign_of_offset}
\end{figure}

\textit{The sign of offset} \cite{nordbeck_computer_1967,haines_point_1994}
method is the simplest point-in-polygon algorithm, specialized for convex
polygons. A point in a convex polygon is shown in (Figure
\ref{fig:sign_of_offset}). For an edge of the polygon, the offset of the point
to the line passing through the edge is calculated. If the offset of the point
has the same sign for all edges of the polygon, the point is inside. Otherwise,
it is outside.

The pseudocode of the implemented algorithm is given in (Algorithm
\ref{alg:offset}).

\begin{algorithm}
    \DontPrintSemicolon
    \SetKwProg{Fn}{Function}{}{}
    \SetKwFunction{Offset}{SignOfOffsetInclusion}
    \Fn(){\Offset}{}{
        \KwData{\\
            \tcp{$V$: Vertices}
            $V \longleftarrow \begin{bmatrix}
                \vect{v_1} & \cdots & \vect{v_n}
            \end{bmatrix}$\;
            \tcp{$Q$: Query Points}
            $Q \longleftarrow \begin{bmatrix}
                \vect{q_1} & \cdots & \vect{q_m}
            \end{bmatrix}$\;
        }
        \KwResult{$IsIn$: Boolean}
        \Begin{
            \tcp{$V'$: Rolled vertices}
            $V' \longleftarrow \begin{bmatrix}
                \vect{v_n}, \vect{v_1} & \cdots & \vect{v_{n-1}}
            \end{bmatrix}$\;
            \tcp{$\Delta$ of successive vertices}
            $\Delta V \longleftarrow (V - V')$\;
            \tcp{LHS \& RHS of the line equation}
            $LHS \longleftarrow Q_y \circ \Delta V_x - Q_x \circ \Delta V_y$\;
            $RHS \longleftarrow V'_y \circ \Delta V_x - V'_x \circ \Delta V _y$\;
            \tcp{Sign test}
            $D \longleftarrow LHS < RHS$\;
            \tcp{Are all same sign}
            $IsIn \longleftarrow Mod_n(\sum_i D_{ij}) = 0$
        }
    }
    \caption{The sign of offset point inclusion test}
    \label{alg:offset}
\end{algorithm}

\subsection{Notation}
\label{notation}

For simplicity and clearance, definitions related to Voronoi tessellations
\cite{okabeSpatialTessellationsConcepts2000} are slightly modified and adapted.

Throughout this paper, only 2-dimensional Euclidean space, $\mathbb{}{R}^2$ is
considered. Boldface denotes a vector, such as $\vect{x} = (x_1, x_2)^T$.
Superscript $^T$ denotes transpose as usual. For a polygon which has $n$
vertices, vertices of the polygon are denoted with additional indexes, such as
$\vect{q_i}, \vect{q_j}$, where $i,j = \{1,\ldots,n\}$ and $i \neq j$ where
edges considered. The set of vertices of the Voronoi polygon is $Q =
\{q_1,\ldots,q_n\}$.

A Voronoi polygon is a convex region, defined by an inner generator point
and some outer generator points such that,

\begin{equation}
    \begin{split}
        V(p_0) = \{ & \vect{x} |\ \|\vect{x} - \vect{p_0}\| \le 
                                  \|\vect{x} - \vect{p_k}\|\\
                    & \forall k \in \{1,\ldots,n\} \}
    \end{split}
    \label{eq:voronoi_polygon}
\end{equation}
where $V(\vect{p_0})$ denotes Voronoi polygon related to the generator point
$\vect{p_0}$.

A generator point $\vect{p_k}$ belongs to the set of generator points $P$ of the
Voronoi polygon. The inner generator point is always indexed as $\vect{p_0}$,
independent of the edge count $n$. For every edge of the Voronoi
polygon there is an outer generator point, so that the set of generator points is
$P = \{\vect{p_0},\vect{p_1},\ldots,\vect{p_n})$.

Edges are equidistant set of points between the inner generator and outer
generators. Precisely,

\begin{equation}
    e_k = \{ \vect{x} |\ \|\vect{x} - \vect{p_0}\| = \|\vect{x} - \vect{p_k}\| \}
    \label{eq:edge}
\end{equation}
where $k \in \{1,\ldots,n\}$. The set of edges of the $V(\vect{p_0})$ can be
denoted as $E = \{e_1,\ldots,e_n\}$.

The whole set of edges is called the boundary, and it is denoted related to the
inner generator point as $\partial V(\vect{p_0})$. Although a Voronoi graph has
multiple polygonal regions, throughout this study, we are only interested in
defining a single Voronoi polygon.

\begin{figure}
    \centering
    \includegraphics[width=1.0\columnwidth]{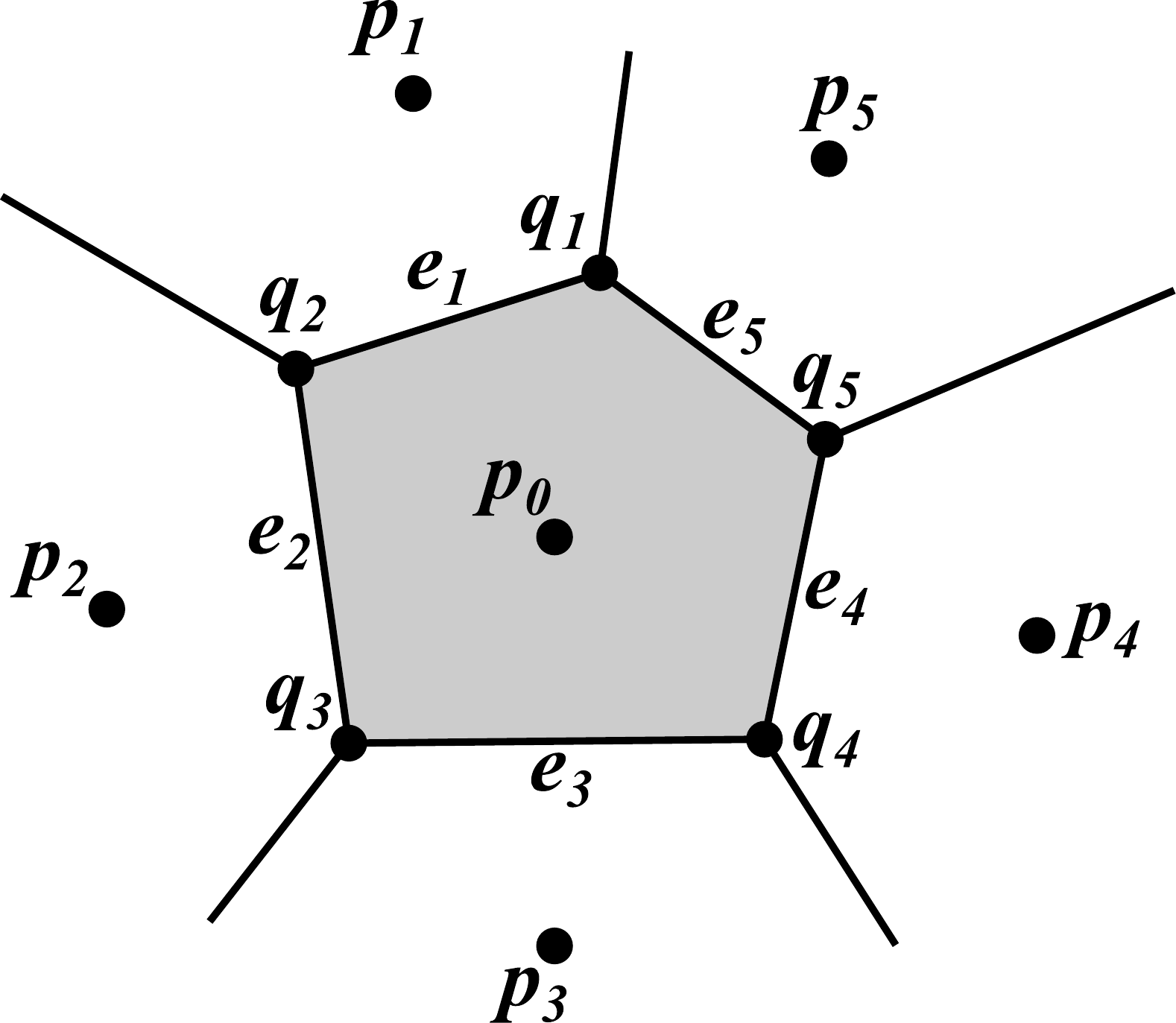}
    \caption{Generators, vertices and edges of a Voronoi polygon}
    \label{fig:gen_pts}
\end{figure}

\section{Conversion of convex polygons to Voronoi polygons}
\label{conversion}

Vertices of a convex polygon ($\vect{q_i}$ in Figure \ref{fig:gen_pts}) can be
taken as the vertices of a Voronoi polygon. Edges of a convex polygon
($\vect{e_k}, \text{where}\ k = \{1,\ldots,5\}$, in Figure \ref{fig:gen_pts})
can be taken as the boundary of a Voronoi polygon.

Because determination of the generator points ($\vect{p_k}$ on Figure
\ref{fig:gen_pts}) is only constrained by $\partial V(\vect{p_0})$, any internal
point can be chosen freely as $\vect{p_0}$. But to distribute the outer generators
homogeneously, and to have a guaranteed point inside, the centroid of the
polygon is used as the inner point. Then, the outer generator points can be found
accordingly.

As shown in (Figure \ref{fig:gen_pts}), placement of generator points determines
both $\partial V(\vect{p_0})$, and the edges going to the infinity. However, our
problem is only constrained on $\partial V(\vect{p_0})$.

The problem of finding $(n+1)$ generator points is constrained on $n$ vertices
of the polygon. So, there is freedom to choose one of the generator points.
Although setting any of the generator points sets all the others, setting the
inner generator is more reasonable; because all the edges are defined depending
upon it.

The centroid of a polygon \cite{bashein_centroid_1994} can be calculated
as follows:

Let $Q$ be a cyclically ordered set of polygon vertices and $\vect{q_i},
\vect{q_j}$ are subsequent vertices accordingly. Summation over $Q$,

\begin{align}
    A = \frac{1}{2}\sum_{Q}^{} \det [\vect{q_i} \vect{q_j}]
    \label{eq:area}\\
    \vect{p_0} = \frac{1}{6 A}\sum_{Q}^{} (\vect{q_i} + \vect{q_j}) \det [\vect{q_i} \vect{q_j}]
    \label{eq:centroid}
\end{align}
gives the area (\ref{eq:area}) and centroid (\ref{eq:centroid}) of the polygon.
The pseudocode of the centroid calculation is given in (Algorithm
\ref{alg:centroid}).

\begin{figure}
    \centering
    \includegraphics[width=0.85\columnwidth]{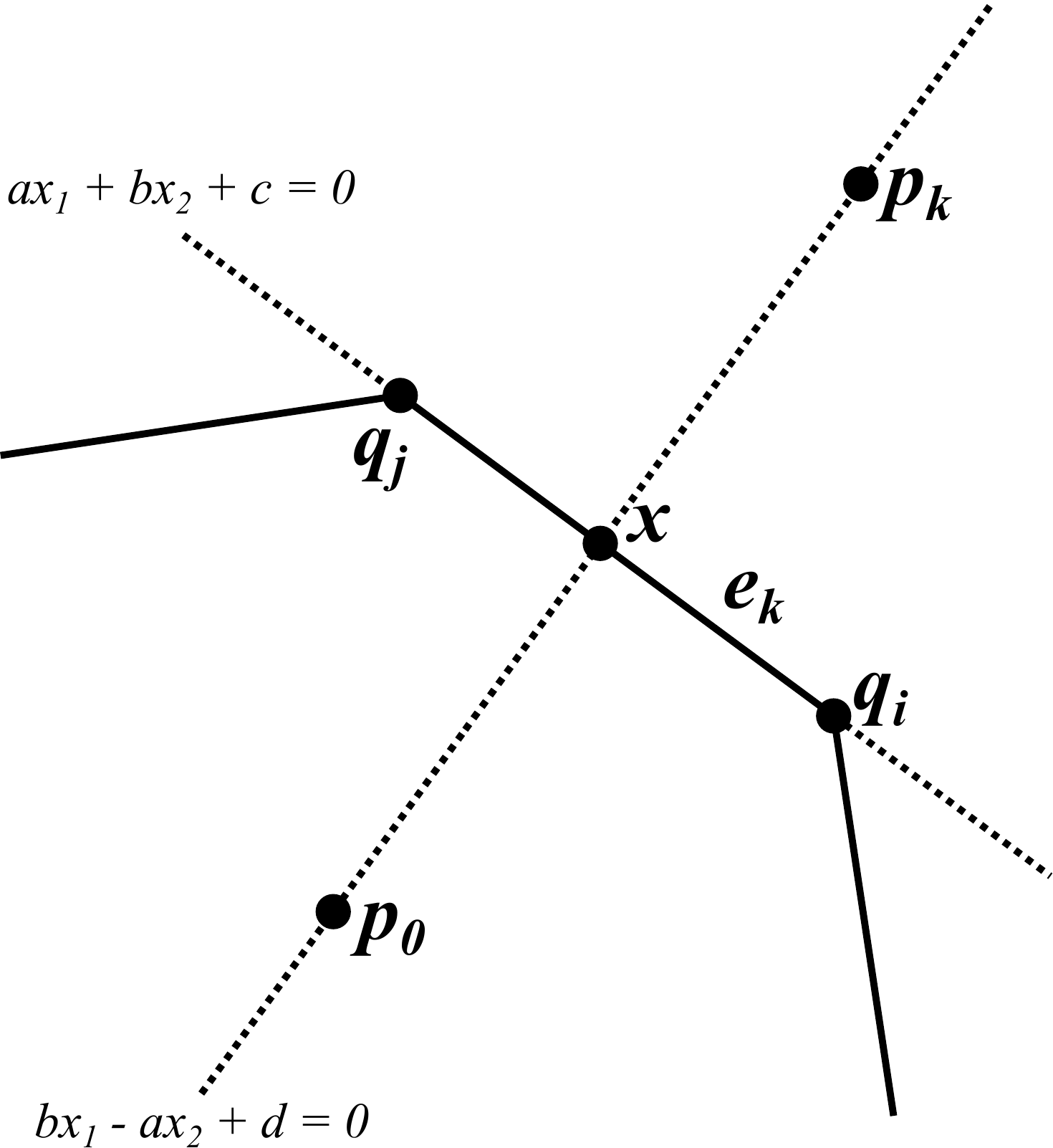}
    \caption{Finding outer generator of an edge}
    \label{fig:reflection}
\end{figure}

\begin{algorithm}
    \SetKwProg{Fn}{Function}{}{}
    \SetKwFunction{Centroid}{CalculateCentroid}
    \DontPrintSemicolon
    \Fn(){\Centroid}{}{
        \KwData{\\
            \tcp{$V$: Vertices}
            $V \longleftarrow \begin{bmatrix}
                \vect{v_1} & \cdots & \vect{v_n}
            \end{bmatrix}$\;
            }
        \KwResult{$\vect{\mu}$: Centroid}
        \Begin{
            \tcp{$V'$: Rolled vertices}
            $V' \longleftarrow \begin{bmatrix}
                \vect{v_n}, \vect{v_1} & \cdots & \vect{v_{n-1}}
            \end{bmatrix}$\;
            \tcp{$A$: Partial areas}
            $A \longleftarrow V'_x \circ V_y - V_x \circ V'_y $\;
            $a \longleftarrow \frac{1}{2} \sum A$ \tcp*{Area}
            $\mu \longleftarrow {\left( (V + V') A \right)} / {(6 a)}$ \tcp*{Centroid}
        }
    }
    \caption{Calculation of the centroid}
    \label{alg:centroid}
\end{algorithm}

For two subsequent vertices $q_i, q_j$ of a polygon, two-point form of
the line equation \cite{weisstein_two-point_nodate} can be written as

\begin{equation}
    (x_2 - q_{i2}) (q_{j1} - q_{i1}) = (x_1 - q_{i1}) (q_{j2} - q_{i2})
    \label{eq:two_point_form}
\end{equation}
where the vertices are $\vect{q_i} = (q_{i1}, q_{i2})^T$ and $\vect{q_j} =
(q_{j1}, q_{j2})^T$.

The standard form equation of the line passing through an edge can be derived
from two-point form equation. As shown in (Figure \ref{fig:reflection}),
$\vect{q_i}$ and $\vect{q_j}$ are two vertices of the edge $e_k$, $\vect{x}$ is
a point on the edge. As defined in (\ref{eq:edge}) $\vect{p_0}$ and $\vect{p_k}$
are two points, equidistant to the $e_k$. The line passing through $\vect{p_0}$
and $\vect{p_k}$ is perpendicular to (\ref{eq:two_point_form}).

Solving $\vect{x}$ for two equations gives
\begin{equation}    
    \vect{x} = \frac{\left (  
        \begin{bmatrix}
            b_k^2 & - a_k b_k\\ 
            - a_k b_k & a_k^2
        \end{bmatrix} \vect{p_0}
        - c_k \begin{bmatrix}
            a_k \\
            b_k
        \end{bmatrix}
        \right )}{{a_k^2 + b_k^2}}
        \label{eq:intersection_x}
\end{equation}
where

\begin{align*}
    a_k = q_{i2} - q_{j2}\\ 
    b_k = q_{j1} - q_{i1}\\ 
    c_k = - (a_k q_{i1} + b_k q_{i2})
\end{align*}

$\vect{p_0}$ and $\vect{p_k}$ are equidistant to $\vect{x}$. Writing this
equation and leaving $\vect{p_k}$ alone on the left hand side gives $\vect{p_k}$
as

\begin{equation}
    \begin{aligned}
        &\vect{p_k} - \vect{x} = \vect{x} - \vect{p_0}\\
        \Rightarrow &\vect{p_k} = 2 \vect{x} - \vect{p_0}
    \end{aligned}
    \label{eq:equidist}
\end{equation}

By substituting (\ref{eq:intersection_x}) into (\ref{eq:equidist}), outer
generator points can be found as

\begin{equation}    
    \vect{p_k} = \frac{\left (  
        \begin{bmatrix}
            b_k^2 - a_k^2 & - 2 a_k b_k\\ 
            - 2 a_k b_k & a_k^2 - b_k^2
        \end{bmatrix} \vect{p_0}
        - 2 c_k \begin{bmatrix}
            a_k \\
            b_k
        \end{bmatrix}
        \right )}{{a_k^2 + b_k^2}}
\end{equation}

The generator calculation procedure is given in (Algorithm
\ref{alg:generators}).

\begin{algorithm}
    \DontPrintSemicolon
    \SetKwProg{Fn}{Function}{}{}
    \SetKwFunction{Generators}{CalculateGenerators}
    \SetKwFunction{Centroid}{CalculateCentroid}
    \Fn(){\Generators}{}{
        \KwData{\\
            \tcp{$V$: Vertices}
            $V \longleftarrow \begin{bmatrix}
                \vect{v_1} & \cdots & \vect{v_n}
            \end{bmatrix}$\;
            }
        \KwResult{$P_{i,j}$: Generators}
        \Begin{
            $P_{i,1} \longleftarrow \Centroid(V)$\;
            \tcp{$V'$: Rolled vertices}
            $V' \longleftarrow \begin{bmatrix}
                \vect{v_n}, \vect{v_1} & \cdots & \vect{v_{n-1}}
            \end{bmatrix}$\;
            $\vect{a} \longleftarrow V_y - V'_y$\;
            $\vect{b} \longleftarrow V'_x - V_x$\;
            $\vect{c} = - (\vect{a} \circ V_x + \vect{b} \circ V_y)$\;
            \tcp{$W$: Weights}
            $W \longleftarrow \begin{bmatrix}
                \vect{b}^2 - \vect{a}^2 & - 2 \vect{a} \vect{b}\\ 
                - 2 \vect{a} \vect{b} & \vect{a}^2 - \vect{b}^2
            \end{bmatrix}$\;
            $\vect{d} \longleftarrow \sum_j W_{ijk} P_{j0}$\;
            $\vect{e} \longleftarrow - 2 \vect{c} \circ \begin{bmatrix}
                \vect{a} \\
                \vect{b}
            \end{bmatrix}$\;
            $P_{i,2:(n+1)} \longleftarrow
                (\vect{d} + \vect{e}) \oslash (\vect{a}^2 + \vect{b}^2)$
        }
    }
    \caption{Calculation of generators}
    \label{alg:generators}
\end{algorithm}

\section{Point inclusion test via generator points}
\label{pip_test}

After the set of generators $P$ has been found, the point inclusion test is
simply testing the condition provided in (\ref{eq:voronoi_polygon}).

\begin{algorithm}
    \DontPrintSemicolon
    \SetKwProg{Fn}{Function}{}{}
    \SetKwFunction{Voronoi}{VoronoiInclusion}
    \SetKwFunction{Generators}{CalculateGenerators}
    \Fn(){\Voronoi}{}{
        \KwData{\\
            \tcp{$V$: Vertices}
            $V \longleftarrow \begin{bmatrix}
                \vect{v_1} & \cdots & \vect{v_n}
            \end{bmatrix}$\;
            \tcp{$Q$: Query Points}
            $Q \longleftarrow \begin{bmatrix}
                \vect{q_1} & \cdots & \vect{q_m}
            \end{bmatrix}$\;
        }
        \KwResult{$IsIn$: Boolean}
        \Begin{
            \tcp{$P$: Generators}
            $P \longleftarrow \Generators(V)$\;
            \tcp{$\Delta$: Differences}
            $\Delta_{ijk} \longleftarrow Q_{i1k} - P_{ij1}$\;
            \tcp{$M$: Metrics}
            $M_{jk} \longleftarrow \sum_{i}^{}{\Delta_{ijk} \Delta_{ijk}}$\;
            $IsIn \longleftarrow M_1 \leq M_j, \forall j \in \{2,\ldots,(n+1)\}$
        }
    }
    \caption{Voronoi point inclusion test}
    \label{alg:voronoi}
\end{algorithm}

The ordinary distance metric for the definition of the Voronoi polygon is
\textit{Euclidean distance} or equivalently \textit{L2 norm}. To test the
inclusion of a random point, its distances to all generators are calculated. If
it is closest to the generator $\vect{p_0}$, it is inside of the polygon.
Otherwise it is outside of the polygon.

Ordinarily, calculating the \textit{L2 norm} of a vector (\ref{eq:L2_norm}) takes squaring, summing and then
square rooting of the vector components.

\begin{equation}
    \|\vect{x}\| = \sqrt{x_1^2 + x_2^2}
    \label{eq:L2_norm}
\end{equation}

However squaring of both sides of (\ref{eq:voronoi_polygon}) does not change the
order of distances, because squaring is a monotonic operation.

\begin{equation}
    V(p_0) = \{ \vect{x} |\ \|\vect{x} - \vect{p_0}\|^2_2 \le 
                            \|\vect{x} - \vect{p_k}\|^2_2 \}
    \label{eq:vpoly_w_sqnorm}
\end{equation}

The square root and the square vanish, when these are applied together. Then
equation (\ref{eq:vpoly_w_sqnorm}) becomes

\begin{equation}
    \begin{aligned}
        V(p_0) = \{ \vect{x} |\ \ & (\vect{x} - \vect{p_0})^T(\vect{x} - \vect{p_0})\\
        \le & (\vect{x} - \vect{p_k})^T(\vect{x} - \vect{p_k}) \}
    \end{aligned}
    \label{eq:vpoly_simpl}
\end{equation}

The derived simplification (\ref{eq:vpoly_simpl}) is an alternate way of
distance comparison. It improves the performance of comparisons and preserves
the order of distances.

The pseudocode of the proposed point inclusion testing algorithm is given in
(Algorithm \ref{alg:voronoi}).

\section{Algorithm analysis}
\label{analysis}

The calculation of the polygon centroid takes $O(n)$ time, when it is done
sequentially. Similarly, the outer generator point calculations have time
complexity of $O(n)$. But considering the Single Instruction Multiple Data
(SIMD) capabilities of modern CPUs, for small sizes of $n$ computations will be
optimized to be done with time complexity of $O(1)$.

For $n$ vertices and $m$ points; $(n+1) m$ distance calculations are done. Then
using distances to the inner centroid as a reference, the number of distance
comparisons to be made is $n m$. Conversion related computations are done
initially, and are independent of the number of processed points. As the number
of points $m$ of the processed points increases, the conversion cost becomes
less effective on the overall computational cost.

In practice, for determination of the status of a point, doing all computations
and comparisons is not always needed. If the point under test is found to be
closer to an outer generator, this breaks the $\forall$ condition of
(\ref{eq:voronoi_polygon}). An early break opportunity arises here for a
sequential implementation of the algorithm.

\section{Experimental results and discussion}
\label{experimental}

\subsection{Correctness}
\label{correctness}

\begin{figure}
    \centering
    \includegraphics[width=1.0\columnwidth]{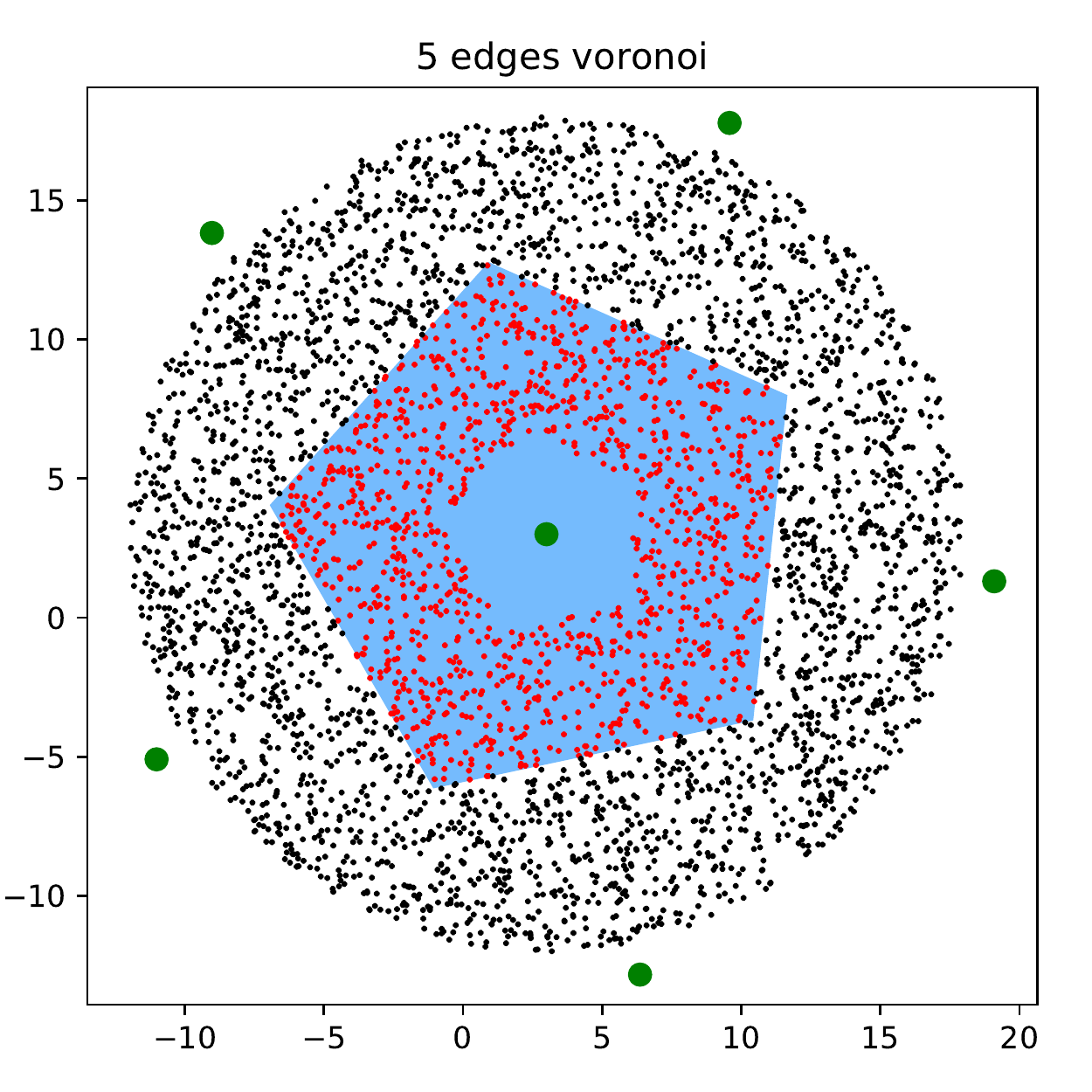}
    \caption{Correctness test of the proposed algorithm}
    \label{fig:correctness}
\end{figure}

To test correctness of the proposed point inclusion algorithm, random test
points are sampled (Figure \ref{fig:correctness}) around the polygon. The set of
generators for the tested polygon are also plotted.

Inclusion test results of \textit{the sign of offset} algorithm are used as the
ground truth. For the same test set, both algorithms gave the
same results. The correctness of the proposed algorithm is proved via this
testing procedure. The correctness of the algorithm can be seen in (Figure
\ref{fig:correctness}) by looking at different coloring of the dots.

\subsection{Performance}
\label{performance}

In order to make a fair comparison, calculations are performed for all vertices,
edges or generators etc. Thus, experimental results reflect theoretical
complexity.

The CPU used for the experimentation is Intel(R) Core(TM) i7-7700, running at
3.60GHz frequency. The system has 32GB of RAM.

For ease of reproducibility, all implementations are done using Python
\cite{van_rossum_python_1995} and related libraries
\cite{van_der_walt_numpy_2011, hunter_matplotlib_2007}. The source code
\cite{rahman_salim_zengin_volimprovoronoi_pip_2021} to reproduce the results is shared.

\begin{figure}
    \centering
    \includegraphics[width=1.0\columnwidth]{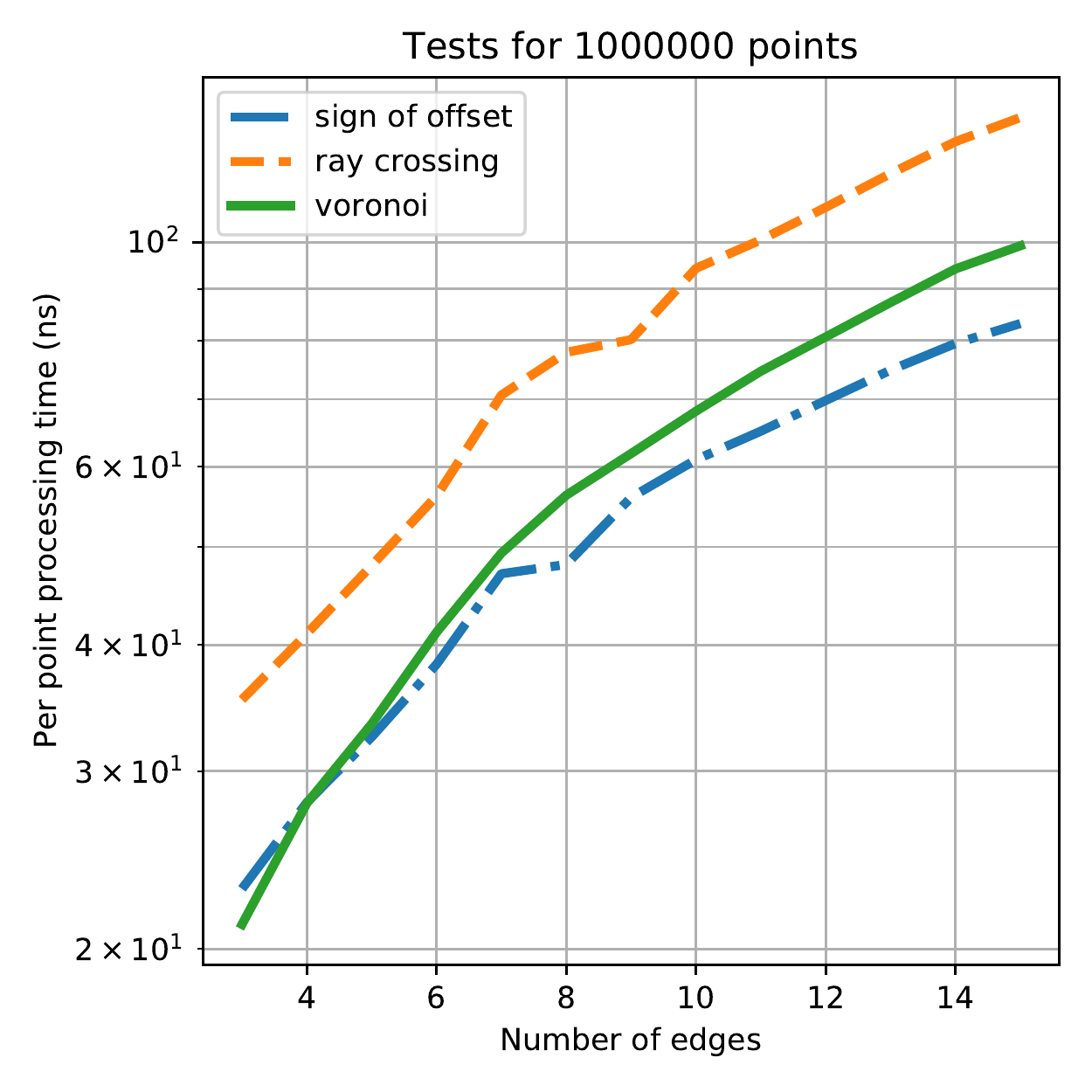}
    \caption{Test results for varying number of edges}
    \label{fig:experimental}
\end{figure}

In order to reduce effect of the runtime overhead, point inclusion tests are
conducted with a point batch size of $1\times10^6$. The number of polygon edges
is changing between $3$ to $15$. As it is illustrated in (Figure
\ref{fig:experimental}), the proposed algorithm gives comparable results with
the reference algorithms.

\section{Conclusion}
\label{conclusion}

A systematic approach to convert a convex polygon to a Voronoi polygon is
developed throughout this work. As a meaningful internal generator point
selection scheme, centroid calculation of a polygon is chosen. The equations,
related to the centroid calculation, are given consecutively. After that the
equations, required to calculate outer generators in relation to the inner
generator and the vertices of the convex polygon, are derived.

In order to demonstrate the advantages of our proposed algorithm, it is
implemented as only vector and matrix operations, without branching. Reference
algorithms are also implemented in a similar way. Certain tests are carried out
to show that, our proposed algorithm not only works properly, but also its
performance is comparable with the reference algorithms.

Conversion of a convex polygon to a Voronoi polygon takes constant time. It only
depends on the number of edges of the polygon. If the geometry is known to be
constant prior to the use, Voronoi equivalent of the convex polygon can be
calculated in advance. Both polygon vertices and generator points can be stored
together in a database with only about $2\times$ of the original storage
capacity needed.

The purpose of this paper is to show that, a precomputed set of test points
based on the Voronoi region idea can be effectively used for testing a point
inside a convex polygon, in a SIMD fashion. The methods developed here can be
extended to nonconvex polygons, and can be applied to prior point-in-polygon
algorithms.



\section*{Acknowledgements}
\label{acknowledgements}

This work was supported by the Turkish Scientific and Technological Research
Council (TUBITAK) under project no. 118E809.

We would like to thank the reviewers for their thoughtful comments and their
constructive remarks.

\newpage

\bibliographystyle{elsarticle-num} 
\bibliography{voronoi_pip.bib}

\end{document}